\documentclass{egblank}
\usepackage{eg2026}
 
\usepackage[T1]{fontenc}
\usepackage{dfadobe}  

\usepackage{cite}  %
\BibtexOrBiblatex
\electronicVersion
\PrintedOrElectronic
\ifpdf \usepackage[pdftex]{graphicx} \pdfcompresslevel=9
\else \usepackage[dvips]{graphicx} \fi

\usepackage{egweblnk} 

\usepackage{enumitem}
\setlist{nosep} %

\usepackage[compact]{titlesec}
\titlespacing{\section}{0pt}{1.5ex}{1ex}
\titlespacing{\subsection}{0pt}{1ex}{0ex}
\titlespacing{\subsubsection}{0pt}{0.5ex}{0ex}

\usepackage{microtype} %

\usepackage{svg}
\usepackage{amsmath}
\usepackage{subcaption}
\usepackage{lipsum}
\usepackage{xcolor}
\usepackage[normalem]{ulem} %
\usepackage{multirow}
\usepackage{siunitx}
\usepackage{subfiles}
\usepackage{comment}
\usepackage[labelfont=bf,textfont=it]{caption}

\definecolor{LivingRoomColor}{HTML}{A6C8E0}
\definecolor{MasterRoomColor}{HTML}{F4B6A6}
\definecolor{KitchenColor}{HTML}{B8D8BA}
\definecolor{BathroomColor}{HTML}{F2C1C1}
\definecolor{DiningRoomColor}{HTML}{C6B7E2}
\definecolor{ChildRoomColor}{HTML}{D3B8AE}
\definecolor{StudyRoomColor}{HTML}{F0CDE3}
\definecolor{SecondRoomColor}{HTML}{CFCFCF}
\definecolor{GuestRoomColor}{HTML}{D8E2A8}
\definecolor{BalconyColor}{HTML}{BFE4E8}
\definecolor{EntranceColor}{HTML}{E1ECF7}
\definecolor{StorageColor}{HTML}{FFE0B5}
\definecolor{WallInColor}{HTML}{CFEBC7}

\newlength{\trimVal}

\newif\ifrevisions
\revisionsfalse

\DeclareRobustCommand{\rev}[1]{%
  \ifrevisions\textcolor{green}{#1}\else#1\fi
}

\DeclareRobustCommand{\del}[1]{%
  \ifrevisions\textcolor{red}{\sout{#1}}\fi
}

\newenvironment{revision}{\begingroup\ifrevisions\color{green}\fi}{\endgroup}

\ifrevisions
    \newenvironment{deletion}{\begingroup\color{red}}{\endgroup}
\else
    \excludecomment{deletion} %
\fi

\title[What a Comfortable World: Ergonomic Principles Guided Apartment Layout Generation]%
      {What a Comfortable World: Ergonomic Principles Guided Apartment Layout Generation}

\begin{revision}

\author[P. Nieciecki \& A. Plocharski \& P. Musialski]
{\parbox{\textwidth}{\centering P. Nieciecki\thanks{Equal contribution}$^{1}$\orcid{0009-0001-7963-9848}, A. Plocharski\footnotemark[1]$^{1,2}$\orcid{0000-0002-7487-8153} and P. Musialski$^{3}$\orcid{0000-0001-6429-8190}
        }
        \\
{\parbox{\textwidth}{\centering $^1$Warsaw University of Technology, Poland\\
         $^2$Akces NCBR, Poland\\
         $^3$New Jersey Institute of Technology, United States of America
       }
}
}

\end{revision}

\begin{document}

\teaser{
 \includegraphics[width=0.9\linewidth, trim={0 60mm 0 0}, clip]{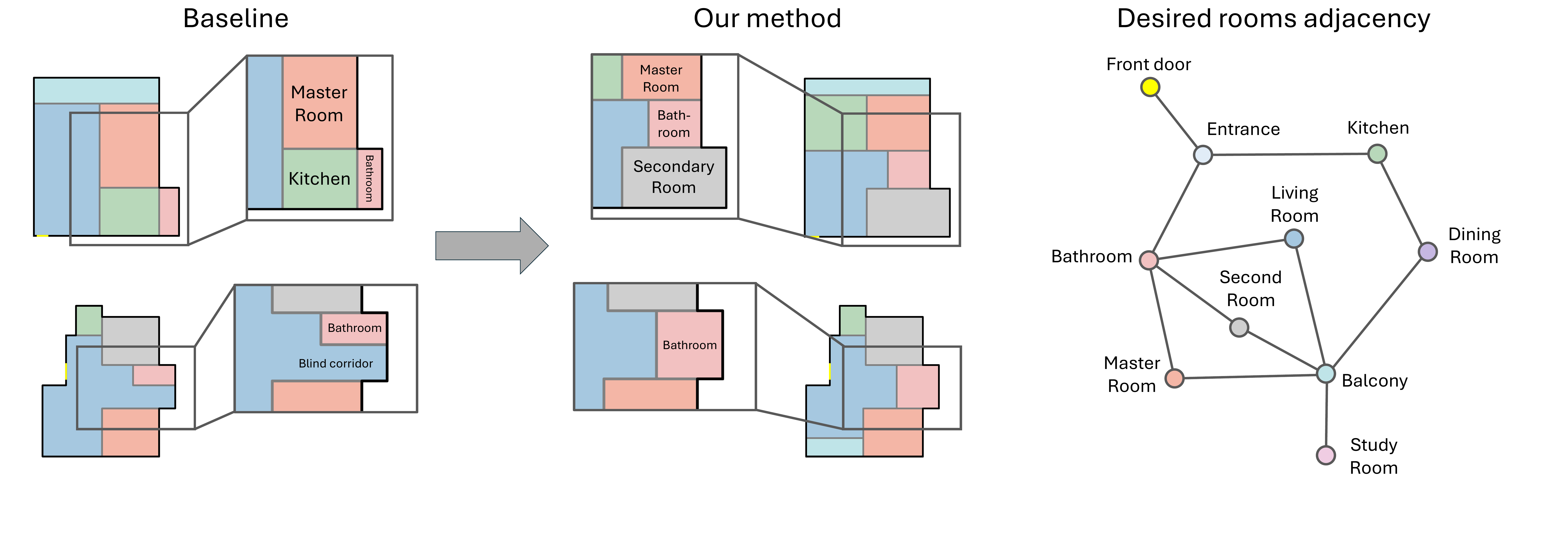}
 \centering
  \caption{The left side illustrates representative examples of ergonomic floor plans generated by the proposed method, with comparison to a baseline method. The right panel displays the adjacency graph defining the desired spatial proximity between specific room pairs.}
\label{fig:teaser}
}

\maketitle

\begin{abstract}
Current data-driven floor plan generation methods often reproduce the ergonomic inefficiencies found in real-world training datasets. To address this, we propose a novel approach that integrates architectural design principles directly into a transformer-based generative process. We formulate differentiable loss functions based on established architectural standards from literature to optimize room adjacency and proximity. By guiding the model with these ergonomic priors during training, our method produces layouts with significantly improved livability metrics. Comparative evaluations show that our approach outperforms baselines in ergonomic compliance while maintaining high structural validity.


\end{abstract}

\section{Introduction}

Automated generation of residential floor plans is a long-standing challenge in computer graphics, aiming to support architects with rapid exploration of design variations. Recent data-driven approaches, particularly those leveraging transformer architectures \del{\cite{gpt2, attention}} \rev{\cite{attention}}, have shown remarkable ability to learn complex spatial distributions from large datasets like RPLAN \cite{RPLAN}. \del{These models treat floor plan generation as a sequence prediction task, generating room boundaries and types token-by-token.}

However, a fundamental limitation of pure data-driven methods is their reliance on the quality of training data. As observed in recent work on furniture arrangement \cite{layoutenhancer}, real-world datasets often contain \del{``imperfect data''---}layouts that are geometrically valid but potentially suboptimal regarding circulation or room placement standards. A model trained naively on such data faithfully reproduces these inefficiencies. While Leimer et al. \cite{layoutenhancer} addressed this by injecting differentiable guidance into the generation of discrete object arrangements, extending this paradigm to \emph{structural floor plan generation} presents a distinct challenge. Unlike moving furniture within a fixed container, generating a floor plan requires optimizing the container itself---defining the topology, adjacency, and tiling of rooms \del{simultaneously}.

In this work, we propose a method to generate residential layouts with improved architectural compliance by integrating domain-specific principles directly into the generative process. We formulate a \del{set of} differentiable loss function\del{s} based on architectural standards \cite{neufert2012architects}\del{, focusing on optimal entrance-to-kitchen proximity and bathroom accessibility}. By incorporating these architectural adjacency \del{and circulation} priors into a GPT-2 based architecture \cite{gpt2} via a dynamic weighting scheme, we guide the model to produce layouts that adhere more closely to design guidelines than the training distribution. 

\noindent\textbf{Our contributions} are: (1) a guideline-derived proximity cost formulation adapted for Manhattan-world room polygons;
\del{(2) a differentiable loss integration compatible with autoregressive token prediction;}
\rev{(2) generalization of differentiable loss integration presented in~\cite{layoutenhancer}}
 and (3) quantitative evidence that our method improves compliance metrics on the RPLAN dataset while maintaining high parsability, albeit with modest trade-offs in area coverage.

\section{Related Work}

Early layout synthesis relied on constraint satisfaction or evolutionary algorithms\rev{~\cite{Michalek2002}}, but the field has shifted toward deep generative models. Graph-based approaches like constraint graphs \cite{para2021generative}, House-GAN++ \cite{nauata2021houseganpp} \rev{and HouseDifusion~\cite{housediffusion}} represent layouts as nodes and edges, ensuring topological validity but often struggling with precise geometry. Most recently, transformer-based autoregressive models have become the state-of-the-art for vector graphics generation. SceneFormer \cite{SceneFormer} and Fa\c{c}AID \cite{plocharski2024facaid} demonstrate that transformers can effectively sequence geometric tokens to generate indoor scenes and building facades, respectively. We build on this sequence-based representation, adapting the GPT-2 architecture \cite{huggingFace} to predict room polygons as coordinate tokens.

Purely generative models often fail to capture high-level functional requirements. To address this, neuro-symbolic methods hybridize neural generation with explicit rule-based constraints. LayoutEnhancer \cite{layoutenhancer} pioneered this for indoor scenes by using differentiable scalar functions to penalize suboptimal furniture arrangements during training. 
This allowed the model to improve upon its own training data. Our work extends this principle from the \emph{interior content} domain (furniture) to the \emph{structural domain} (floor plans). 
We adapt the loss formulation to operate on polygon vertices and adjacencies, ensuring that the generated structures align with established architectural circulation guidelines \cite{neufert2012architects}.

\section{Ergonomic cost}\label{sec:ergo cost}

Using domain-specific literature \cite{neufert2012architects} we determined that minimizing the following distances have a positive impact on the overall comfort of apartment use:
    (i) the distance from the entrance room to the front door;
    (ii) the distances from the entrance room, living room, master room and second room to the bathroom;
    (iii) the distances from the entrance room and dining room to the kitchen;
\begin{revision}
    (iv) balcony adjacency to kitchen, dining room, living room, master room, secondary room, or study room.
\end{revision}

\del{Additionally, it is desirable for the balcony to be adjacent to one of the following rooms: kitchen, dining room, living room, master room, secondary room, or study room.}

Taking the above aspects into account, we define a set of room-type-specific cost functions to quantify layout ergonomics.

\noindent
\textbf{Entrance cost.} For each entrance room polygon $r\in R_\mathrm{entrance}$, the entrance cost is defined as
$$
E_\mathrm{entrance}(r) = \mathit{dist}(r, d),
$$
\del{where $d$ is the front door and $\mathit{dist}$ denotes distance in meters.}
\rev{where $d$ denotes the front door (represented as a line segment) and $\mathit{dist}$ is the Euclidean distance in meters.}

\noindent
\textbf{Kitchen cost.} For kitchens, we first pair each entrance and dining room with the nearest kitchen present in the floor plan. The cost associated with a given kitchen is then computed as the average distance to all entrances and dining areas assigned to it:
$$
E_\mathrm{kitchen}(r) = \frac{1}{|A(r)|} \sum_{r'\in A(r)} \mathit{dist}(r, r'),
$$
where $r\in R_\mathrm{kitchen}$ is a kitchen and $A(r)$ is a set of rooms assigned to kitchen $r$.

\begin{deletion}
    \noindent
    \textbf{Bathroom cost.} The bathroom cost is defined in an analogous manner, but for each entrance room, living room, master room and second room we assign the nearest bathroom:
    $$
    E_\mathrm{bathroom}(r) = \frac{1}{|A(r)|} \sum_{r'\in A(r)} \mathit{dist(r, r')},
    $$
    where $r\in R_\mathrm{bathroom}$ is a bathroom.
\end{deletion}

\begin{revision}
    \noindent
    \textbf{Bathroom cost.} The cost of each bathroom $r\in R_\mathrm{bathroom}$ is defined in an analogous manner, but for each entrance room, living room, master room and second room we assign the nearest bathroom:
    $$
    E_\mathrm{bathroom}(r) = \frac{1}{|A(r)|} \sum_{r'\in A(r)} \mathit{dist(r, r')},
    $$
\end{revision}

\begin{deletion}
    \noindent
    \textbf{Balcony cost.} For balconies, the cost is defined as the minimum distance to any of the preferred adjacent room types:
    $$
    E_\mathrm{balcony}(r) = \mathrm{min}_{r' \in R'} \mathit{dist}(r, r'),
    $$
    where $R'$ is a set of desired neighbors for a balcony -- \rev{see Fig.~\ref{fig:teaser}}.\del{ -- kitchens, dining rooms, living rooms, master rooms, secondary rooms and study rooms.}
\end{deletion}

\begin{revision}
    \noindent
    \textbf{Balcony cost.} The cost is defined as the minimum distance to any room from the set of preferred adjacent room types ($R'$):
    $$
    E_\mathrm{balcony}(r) = \mathrm{min}_{r' \in R'} \mathit{dist}(r, r').
    $$
\end{revision}

Overall ergonomic cost is computed as the mean cost over all rooms of applicable types:
\begin{equation} \label{eq: ergonomic cost}
    E = \frac{1}{|R^*|} \sum_{r \in R^*} E_{a(r)}(r),
\end{equation}
with $R^* = R_\mathrm{entrance} \cup R_\mathrm{kitchen} \cup R_\mathrm{bathroom} \cup R_\mathrm{balcony}$ and $a(r)$ is the type of room $r$.

\section{Floor plans generation}

Following trends in related work, we employ the GPT-2 model \cite{gpt2} (specifically the implementation included in the Hugging Face library \cite{huggingFace}). \del{which is based on the transformer architecture \cite{attention}. In this section, we describe the application of this model to the task of floor plan design, with a particular emphasis on the non-trivial incorporation of expert domain knowledge.}

\subsection{Floor plan representation}

We encode each floor plan as a token sequence $S=(b,d,r_1,\ldots,r_n)$, where $b$ is the boundary, $d$ the front door placement, and $r_i$ the $i$-th room. Boundary and door appear first; room order may vary.

Each segment starts with a type token ($s_b$ for boundary, $s_d$ for door, $s_r^t$ for a room of type $t$) followed by quantized vertex coordinates $(x,y)$ at fixed resolution:
\begin{align*}
b &= (s_b, x_{b1}, y_{b1}, x_{b2}, y_{b2}, \ldots, x_{bk}, y_{bk}) \\
d &= (s_d, x_{d1}, y_{d1}, x_{d2}, y_{d2}) \\
r_i &= (s_i^t, x_{i1}, y_{i1}, x_{i2}, y_{i2}, \ldots, x_{il}, y_{il})
\end{align*}

Following~\cite{SceneFormer, para2021generative, layoutenhancer, plocharski2024facaid}, we augment each token with an xy-index and a vertex-index. The xy-index is 1 for x-coordinates, 2 for y-coordinates, and 0 otherwise. The vertex-index is 0 for start tokens and counts coordinate pairs (1 for the first vertex, 2 for the second, etc.). Each index has its own learned embedding, added to the token and positional embeddings.

\subsection{Ergonomic loss}

To inject expert knowledge into the model, we introduce a custom loss function using the ergonomic cost terms described in Section~\ref{sec:ergo cost}.

First, we define a differentiable distance metric between rooms. For room polygons $r$ and $s$ let $V_r = \{p^r_1, p^r_2, \ldots p^r_{n_r} \}$ and $V_s = \{p^s_1, p^s_2, \ldots p^s_{n_s} \}$ be the sets of their vertices. Then the distance metric $D$ is defined as:
$$
D(r,s) = \langle e, \text{softmin}(\beta \cdot e) \rangle, \:\:\:\text{where} \:\:\:e_{ij} = \|p^r_i - p^s_j\|_2.
$$
Here, $\beta$ is a temperature parameter that determines the hardness of the softmin function (our experiments use $\beta = 10$).

Similarly to ergonomic cost, ergonomic loss is defined as a combinations of room-type-specific losses. 
For each entrance room it is equal to the differentiable distance between the entrance room polygon and the front door polygon.
$$
L_{entrance}(r) = D(r, d),
$$
where $d$ is the front door polygon. The mean value over all entrance rooms in a floor plan forms the full entrance room loss term:
$$
L_{entrances} = \frac{1}{|R_{entrance}|} \sum_{r \in R_{entrance}} L_{entrance}(r).
$$

For kitchens, each entrance room and dining room distance is calculated to each kitchen. Then a differentiable minimum function is used to determine the distance from the nearest kitchen. The kitchen loss term is a mean value of these distances. Let $r_i \in R_{entrance} \cup R_{dining}$, $r_j \in R_{kitchen}$ and $\Delta_{ij} = D(r_i, r_j)$, then:
\begin{deletion}
    $$
    L_{kitchen}(r_i) = \langle \Delta_{ij}, \text{softmin}(\beta \cdot \Delta_{ij}) \rangle,
    $$
    $$
    L_{kitchens} = \frac{1}{|R_{entrance} \ \cup R_{dining}|} \sum_{r_i \in R_{entrance} \ \cup R_{dining}} L_{kitchen}(r_i).
    $$
\end{deletion}
\begin{revision}
    \begin{gather*}
    L_{kitchen}(r_i) = \langle \Delta_{ij}, \text{softmin}(\beta \cdot \Delta_{ij}) \rangle, \\
    L_{kitchens} = \frac{1}{|R_{entrance} \ \cup R_{dining}|} \sum_{r_i \in R_{entrance} \ \cup R_{dining}} L_{kitchen}(r_i).
    \end{gather*}
\end{revision}

Loss for bathrooms is defined in an analogical way, but using the proximity of entrances, living rooms, master rooms and second rooms\del{:} \rev{. Let $R'' = R_{entrance} \cup R_{living} \cup R_{master} \cup R_{second}$, then:} 
$$
L_{bathrooms} = \frac{1}{|R''|} \sum_{r_i \in R''} L_{bathroom}(r_i).
$$
\del{where $R'' = R_{entrance} \cup R_{living} \cup R_{master} \cup R_{second}$.}

Balcony loss is similar, but here we are taking a differentiable minimum of proximate distance losses:
$$
L_{balconies} = \langle L_{balcony}(r_i), \text{softmin}(\beta \cdot L_{balcony}(r_i) \rangle,\:\:\: r_i \in R'.
$$

Final ergonomic loss is a mean \del{value} of all possible to calculate losses:
$$
L_E = \frac{\sum_a \delta_a L_a}{\sum_a \delta_a},
$$
with $a = \{ entrances, \ kitchens, \ bathrooms, \ balconies \}$ and $\delta_a = 1$ if the\del{ corresponding} loss is applicable to the \del{particular} floor plan and $\delta_a = 0$ otherwise.

\subsection{Ergonomic loss usage}
\label{sec:loss_usage}

We integrate our ergonomic loss into the training process using a methodology similar to one presented in \cite{layoutenhancer}.
The ergonomic loss is calculated using ground-truth sequences where a single token is replaced by a value $\bar{v}$ derived from the model's predicted probability distribution.

Token $\bar{v}$ is equal to the expected value in a small window around the most likely value of the token
$$
\bar{v} = \frac{\sum_j\mathcal{N}(v_j|\hat{v}, \sigma) P(v_j)v_j}{\sum_j\mathcal{N}(v_j|\hat{v}, \sigma) P(v_j)},
$$
with $\mathcal{N}(x|\hat{v}, \sigma)$ being the normal distribution centered at $\hat{v}$ with standard deviation $\sigma$. $\hat{v}$ is equal to the most probable token and $P(v)$ is a estimated by the model. During our experiments $\sigma$ was equal to $1/\rho$, where $\rho$ is the resolution of floor plans quantization\del{ (number of tokens, which represents x- or y-coordinates)}.

Since the ergonomic loss is differentiable w.r.t. the room polygon vertices, the loss value for a given predicted token $\bar{v}$ is computed only when both $\hat{v}$ and its corresponding ground-truth element represent x- or y-coordinate of a room polygon vertex.

During the learning process we are combining the standard cross-entropy loss $L_C$ with our ergonomic loss.
$$
L = (1-\alpha) \cdot L_C + \alpha \cdot L_E,
$$
where $\alpha$ is \del{equal to} the value of our ergonomic loss of the ground-truth floor plan divided by a scaling factor $\gamma$ and clamped to $[0; 1]$ range. \del{During our experiments $\gamma$ was equal to 30 and it was derived from the distribution of ergonomic losses calculated on test samples from the dataset. This definition ensures that ergonomic loss has stronger influence when training on non-ergonomic samples from the dataset. This makes the model lean more on expert knowledge for data points that don't align with ergonomic guidelines.}
\rev{During our experiments, $\gamma$ was set to 30 based on ergonomic loss distribution from test samples. This definition increases the influence of ergonomic loss on non-ergonomic data, encouraging the model to rely more on expert knowledge for such cases.}

\rev{\documentclass[ShortPaper]{subfiles}

\begin{document}

\begin{figure*}[t]
\centering

\renewcommand{\arraystretch}{0.9}
\setlength{\trimVal}{20pt}

\begin{minipage}{0.31\textwidth}
    \centering
    \setlength{\tabcolsep}{0pt}
    \begin{tabular}{@{}c@{}c@{}c@{}}
        \multicolumn{3}{c}{Inaccessible bathrooms} \\
        \textbf{Our} & \textbf{Baseline} & \textbf{RPLAN} \\

        \includegraphics[width=0.33\linewidth, trim={\the\trimVal} {\the\trimVal} {\the\trimVal} {\the\trimVal}, clip]{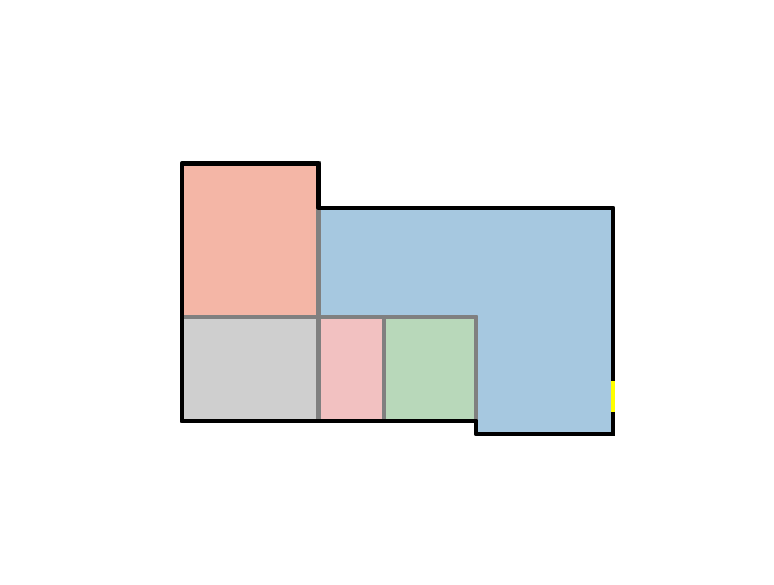} &
        \includegraphics[width=0.33\linewidth, trim={\the\trimVal} {\the\trimVal} {\the\trimVal} {\the\trimVal}, clip]{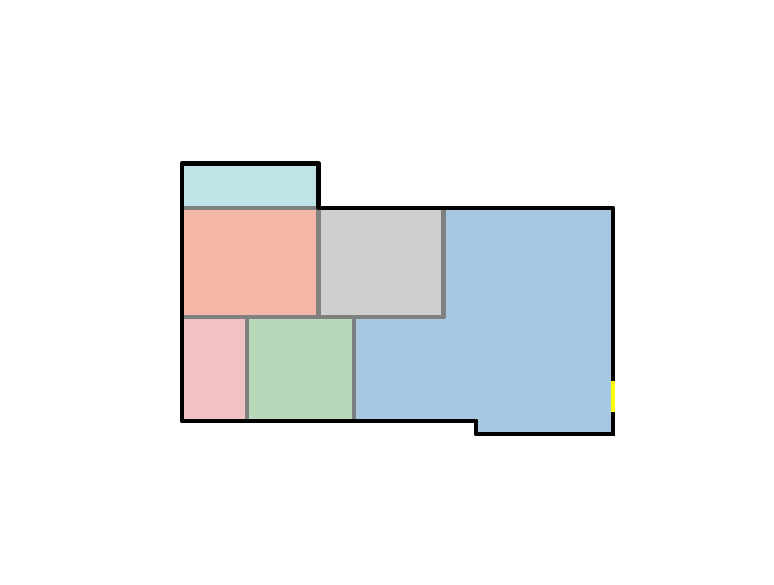} &
        \includegraphics[width=0.33\linewidth, trim={\the\trimVal} {\the\trimVal} {\the\trimVal} {\the\trimVal}, clip]{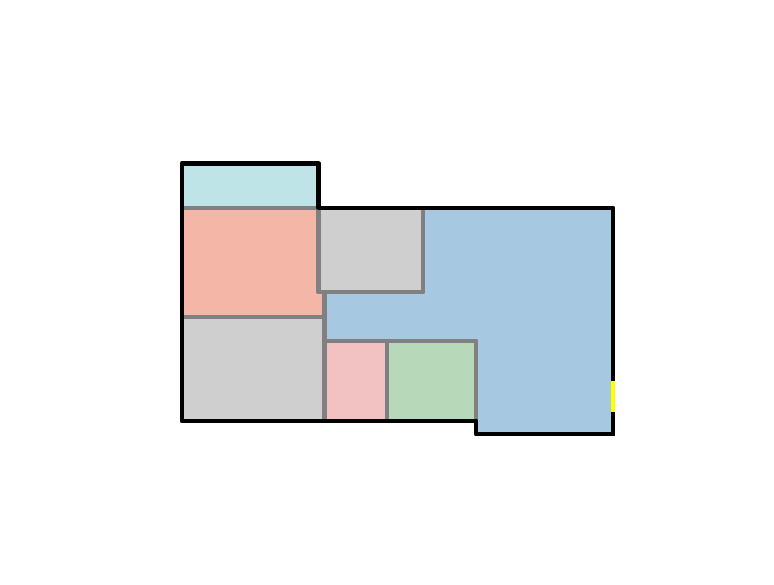} \\
        \includegraphics[width=0.33\linewidth, trim={\the\trimVal} {\the\trimVal} {\the\trimVal} {\the\trimVal}, clip]{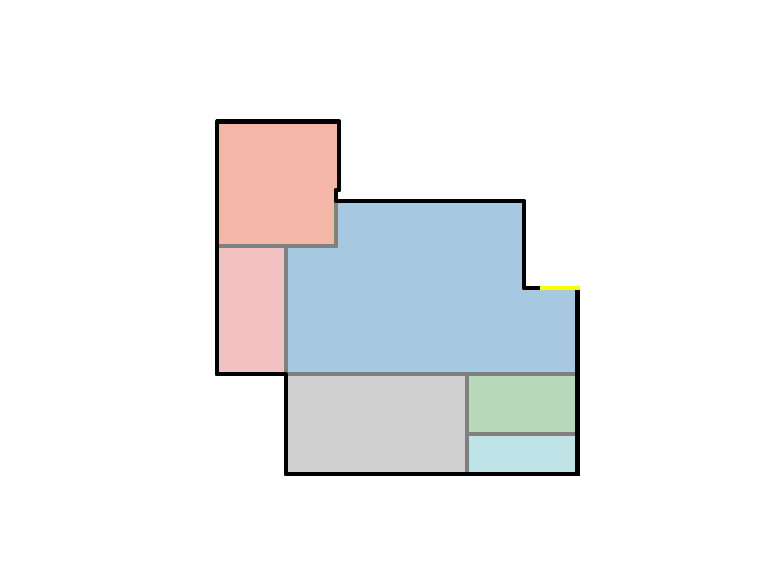} &
        \includegraphics[width=0.33\linewidth, trim={\the\trimVal} {\the\trimVal} {\the\trimVal} {\the\trimVal}, clip]{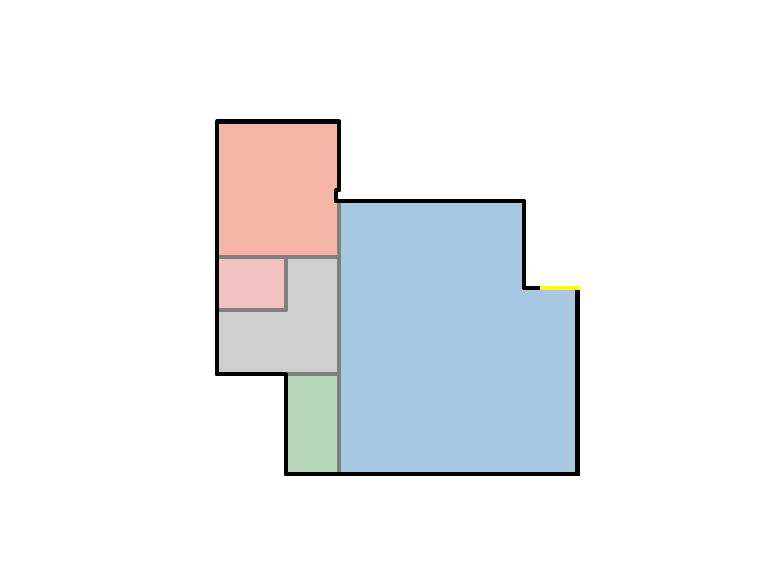} &
        \includegraphics[width=0.33\linewidth, trim={\the\trimVal} {\the\trimVal} {\the\trimVal} {\the\trimVal}, clip]{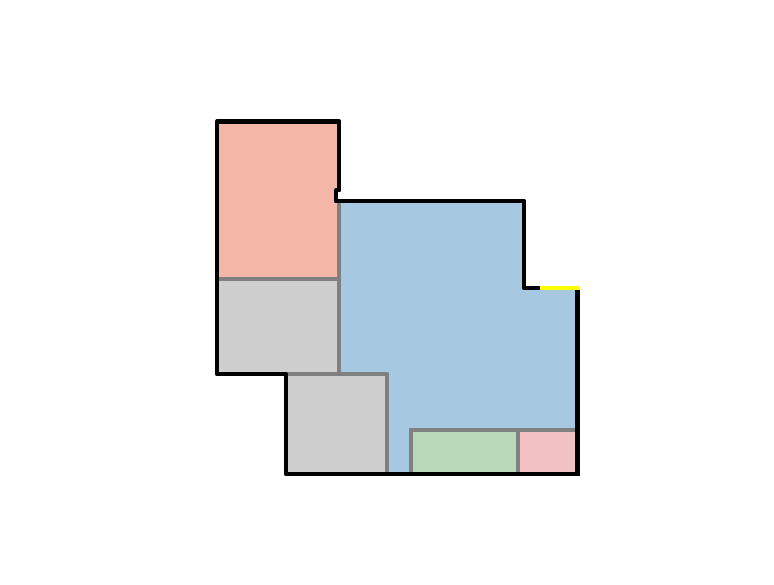} \\

        \includegraphics[width=0.33\linewidth, trim={\the\trimVal} {\the\trimVal} {\the\trimVal} {\the\trimVal}, clip]{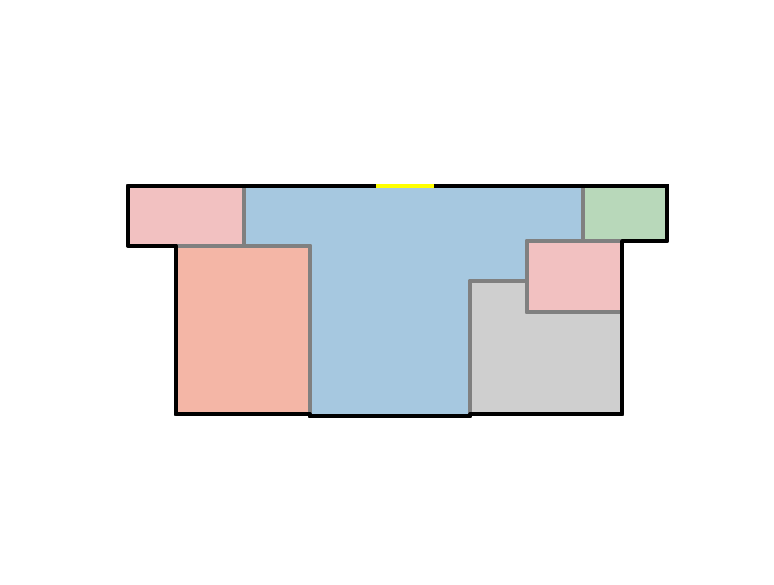} &
        \includegraphics[width=0.33\linewidth, trim={\the\trimVal} {\the\trimVal} {\the\trimVal} {\the\trimVal}, clip]{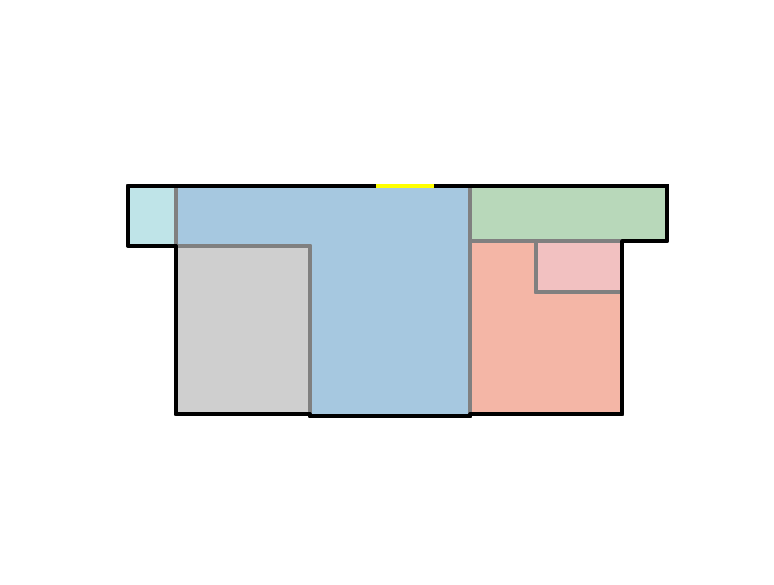} &
        \includegraphics[width=0.33\linewidth, trim={\the\trimVal} {\the\trimVal} {\the\trimVal} {\the\trimVal}, clip]{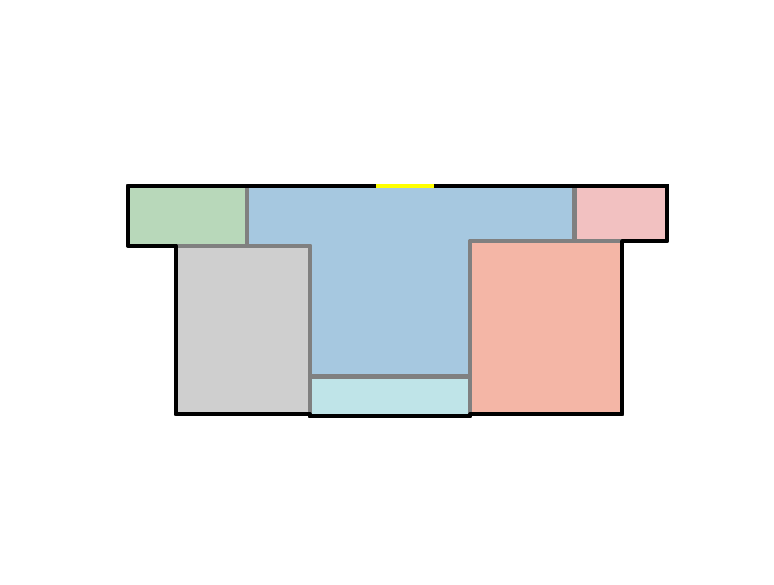} \\

    \end{tabular}
\end{minipage}
\hfill
\begin{minipage}{0.31\textwidth}
    \centering
    \setlength{\tabcolsep}{0pt}
    \begin{tabular}{@{}c@{}c@{}c@{}}
        \multicolumn{3}{c}{Blind corridors} \\
        \textbf{Our} & \textbf{Baseline} & \textbf{RPLAN} \\

        \includegraphics[width=0.33\linewidth, trim={\the\trimVal} {\the\trimVal} {\the\trimVal} {\the\trimVal}, clip]{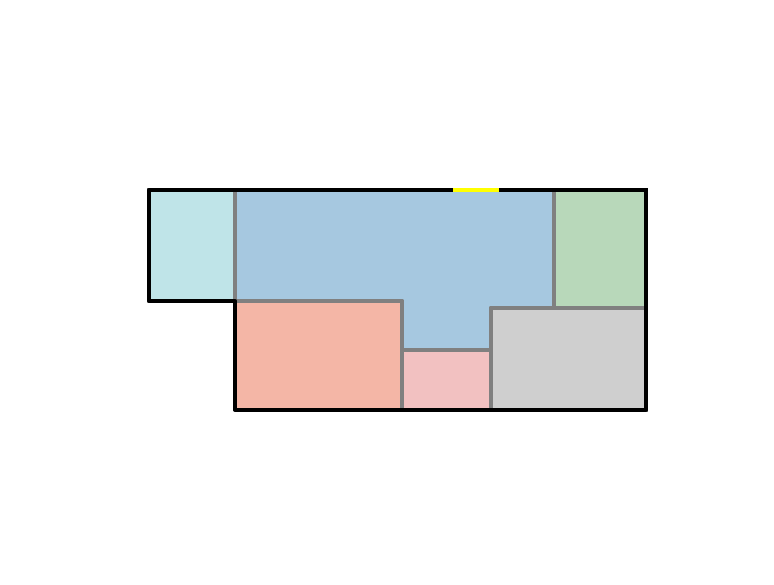} &
        \includegraphics[width=0.33\linewidth, trim={\the\trimVal} {\the\trimVal} {\the\trimVal} {\the\trimVal}, clip]{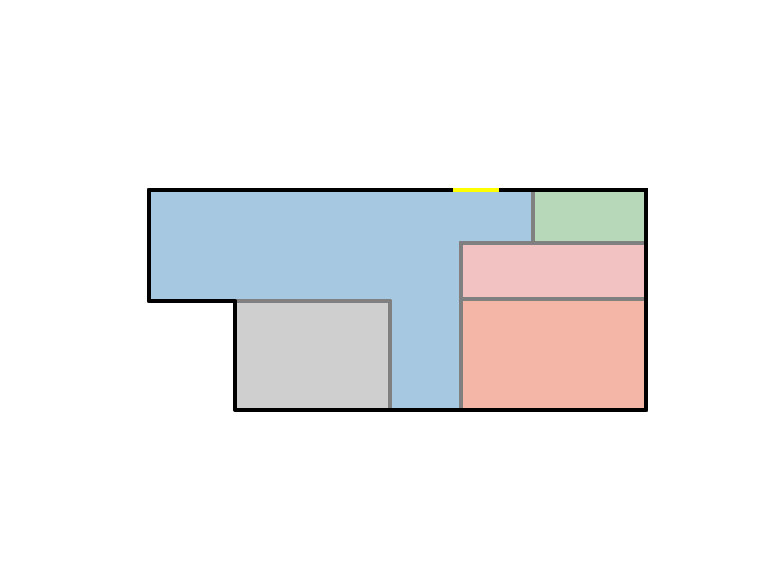} &
        \includegraphics[width=0.33\linewidth, trim={\the\trimVal} {\the\trimVal} {\the\trimVal} {\the\trimVal}, clip]{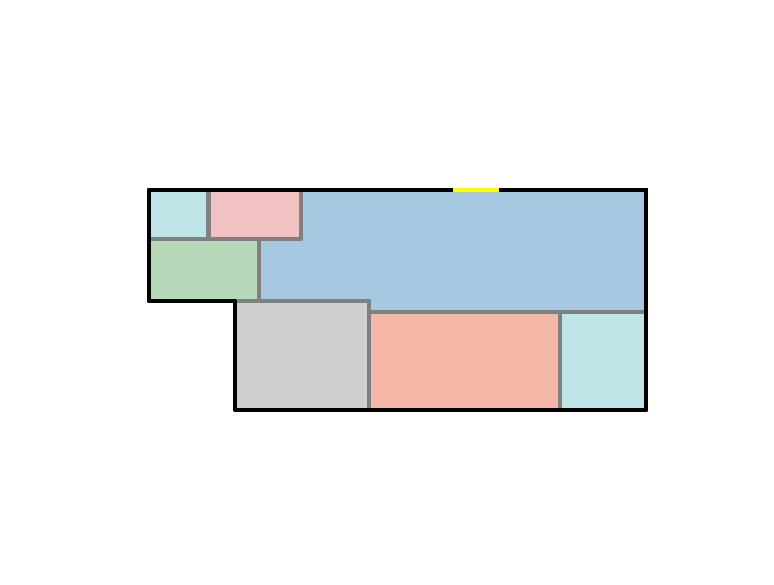} \\

        \includegraphics[width=0.33\linewidth, trim={\the\trimVal} {\the\trimVal} {\the\trimVal} {\the\trimVal}, clip]{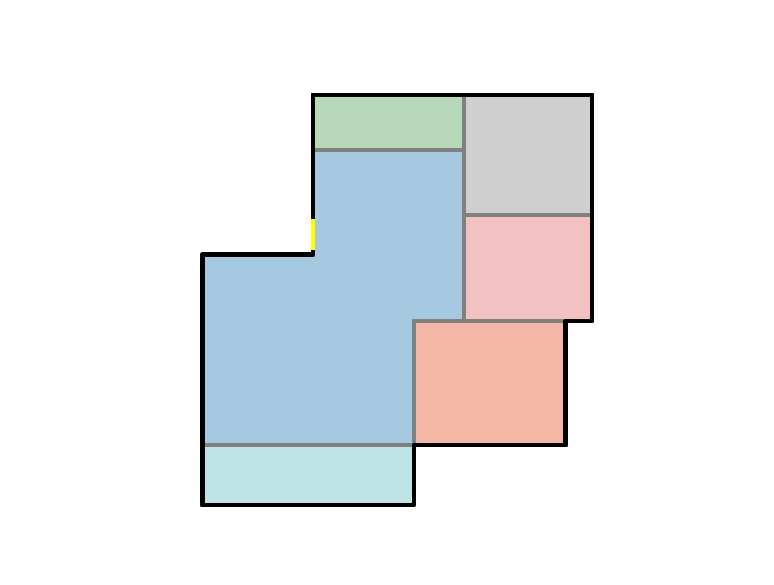} &
        \includegraphics[width=0.33\linewidth, trim={\the\trimVal} {\the\trimVal} {\the\trimVal} {\the\trimVal}, clip]{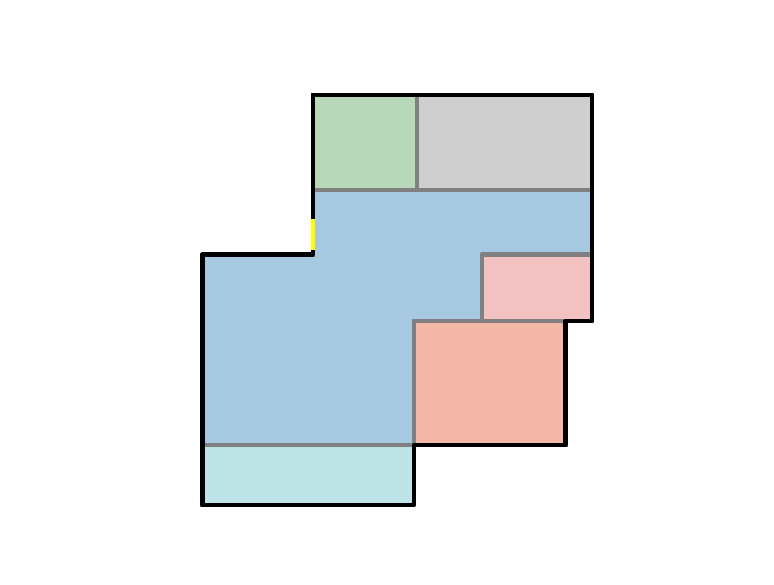} &
        \includegraphics[width=0.33\linewidth, trim={\the\trimVal} {\the\trimVal} {\the\trimVal} {\the\trimVal}, clip]{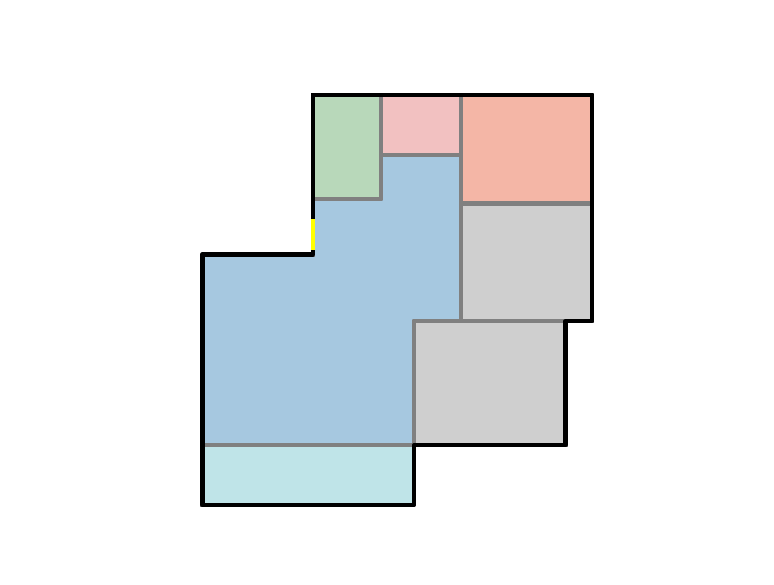} \\

        \includegraphics[width=0.33\linewidth, trim={\the\trimVal} {\the\trimVal} {\the\trimVal} {\the\trimVal}, clip]{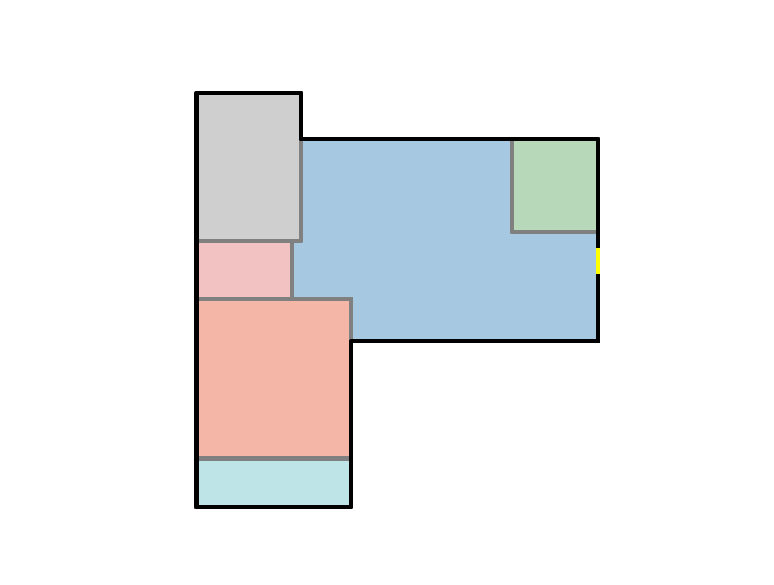} &
        \includegraphics[width=0.33\linewidth, trim={\the\trimVal} {\the\trimVal} {\the\trimVal} {\the\trimVal}, clip]{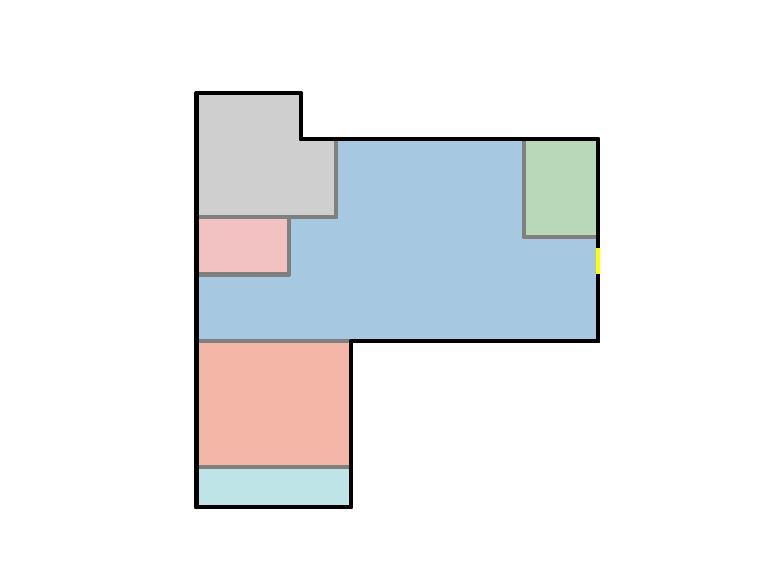} &
        \includegraphics[width=0.33\linewidth, trim={\the\trimVal} {\the\trimVal} {\the\trimVal} {\the\trimVal}, clip]{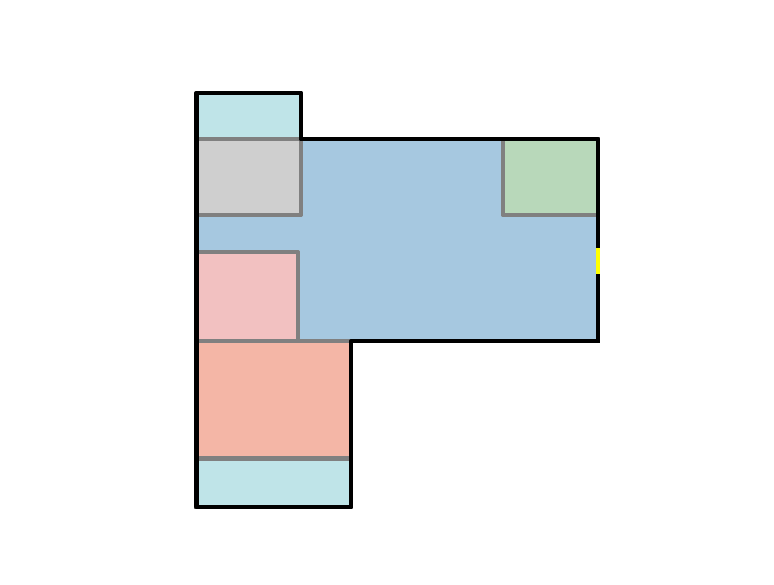} \\

    \end{tabular}
\end{minipage}
\hfill
\begin{minipage}{0.31\textwidth}
    \centering
    \setlength{\tabcolsep}{0pt}
    \begin{tabular}{@{}c@{}c@{}c@{}}
        \multicolumn{3}{c}{Better room arrangement} \\
        \textbf{Our} & \textbf{Baseline} & \textbf{RPLAN} \\

        \includegraphics[width=0.33\linewidth, trim={\the\trimVal} {\the\trimVal} {\the\trimVal} {\the\trimVal}, clip]{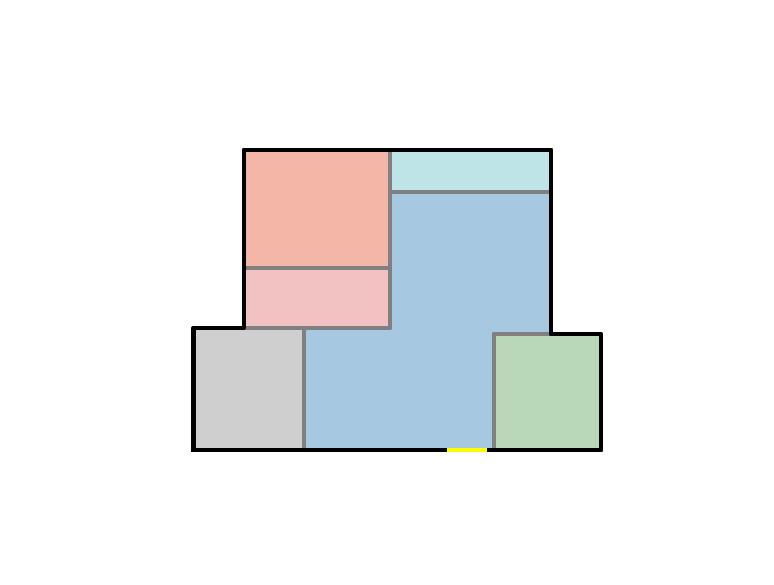} &
        \includegraphics[width=0.33\linewidth, trim={\the\trimVal} {\the\trimVal} {\the\trimVal} {\the\trimVal}, clip]{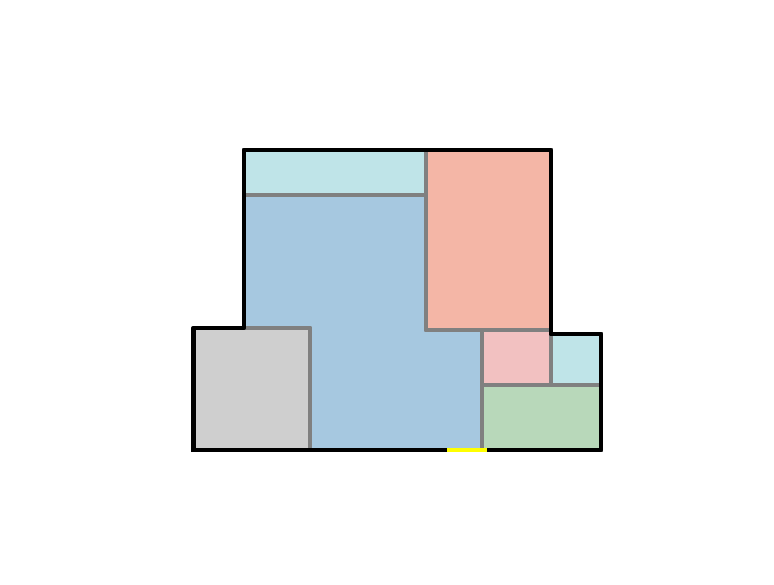} &
        \includegraphics[width=0.33\linewidth, trim={\the\trimVal} {\the\trimVal} {\the\trimVal} {\the\trimVal}, clip]{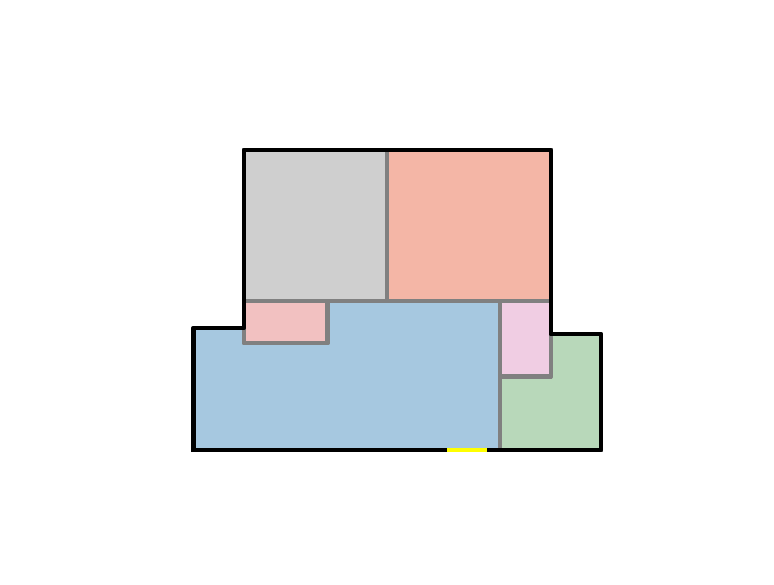} \\

        \includegraphics[width=0.33\linewidth, trim={\the\trimVal} {\the\trimVal} {\the\trimVal} {\the\trimVal}, clip]{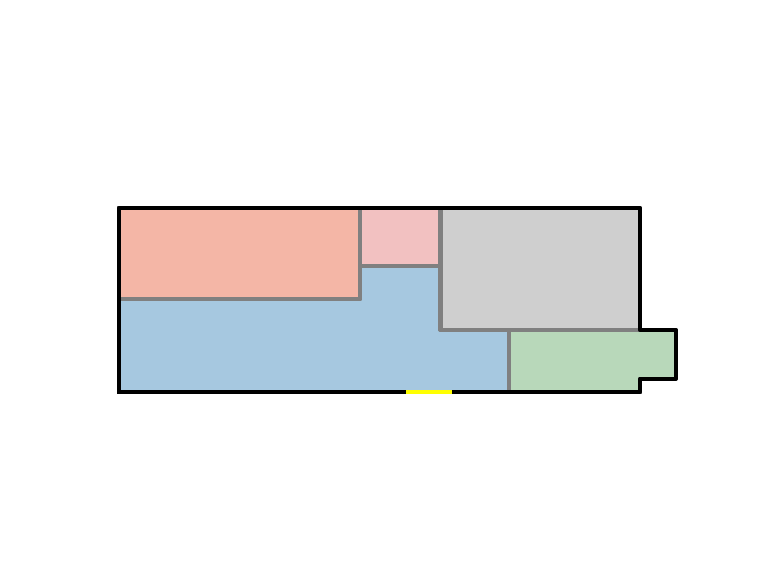} &
        \includegraphics[width=0.33\linewidth, trim={\the\trimVal} {\the\trimVal} {\the\trimVal} {\the\trimVal}, clip]{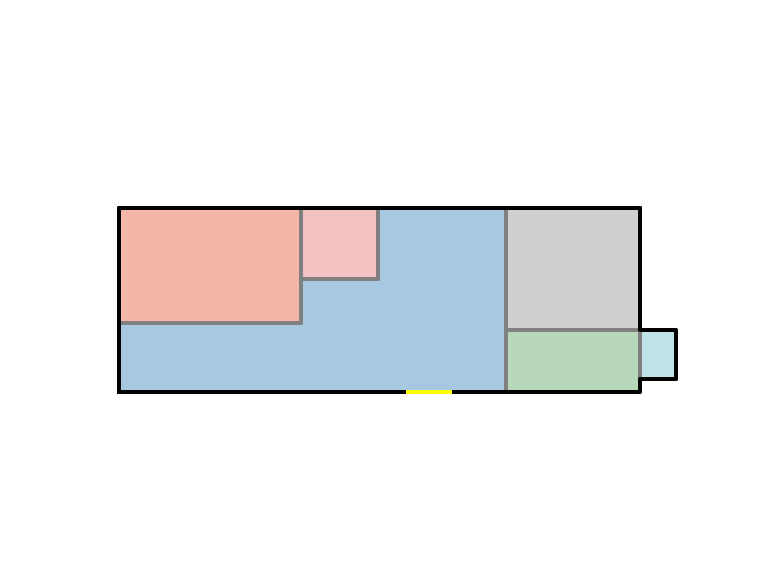} &
        \includegraphics[width=0.33\linewidth, trim={\the\trimVal} {\the\trimVal} {\the\trimVal} {\the\trimVal}, clip]{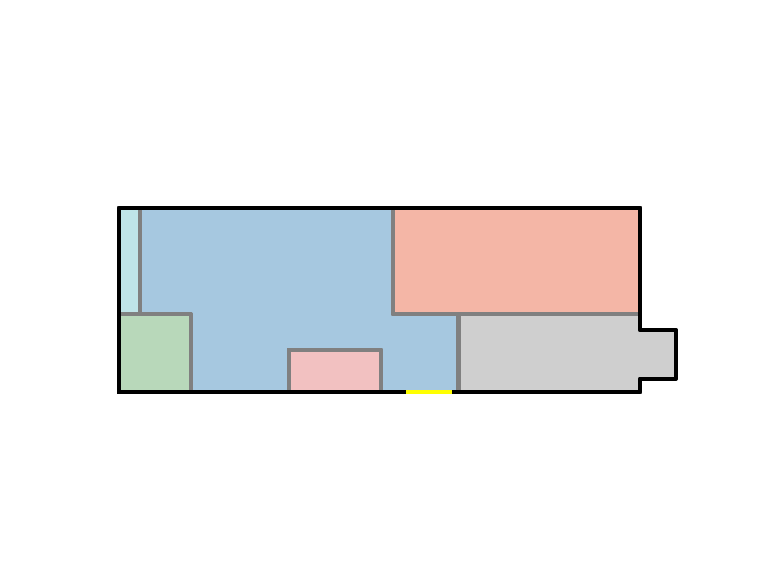} \\

        \includegraphics[width=0.33\linewidth, trim={\the\trimVal} {\the\trimVal} {\the\trimVal} {\the\trimVal}, clip]{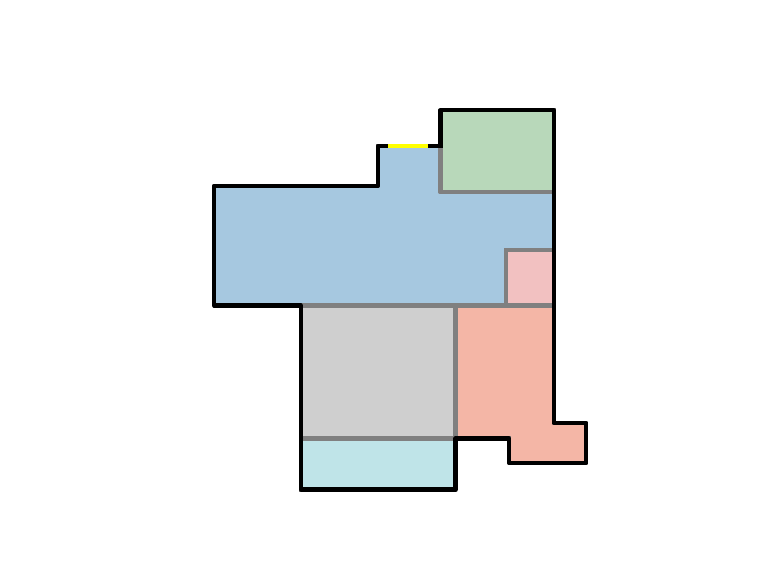} &
        \includegraphics[width=0.33\linewidth, trim={\the\trimVal} {\the\trimVal} {\the\trimVal} {\the\trimVal}, clip]{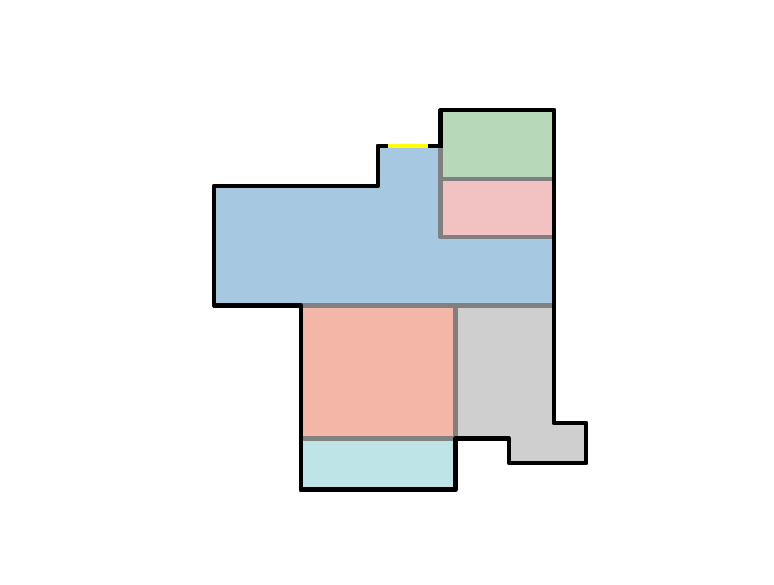} &
        \includegraphics[width=0.33\linewidth, trim={\the\trimVal} {\the\trimVal} {\the\trimVal} {\the\trimVal}, clip]{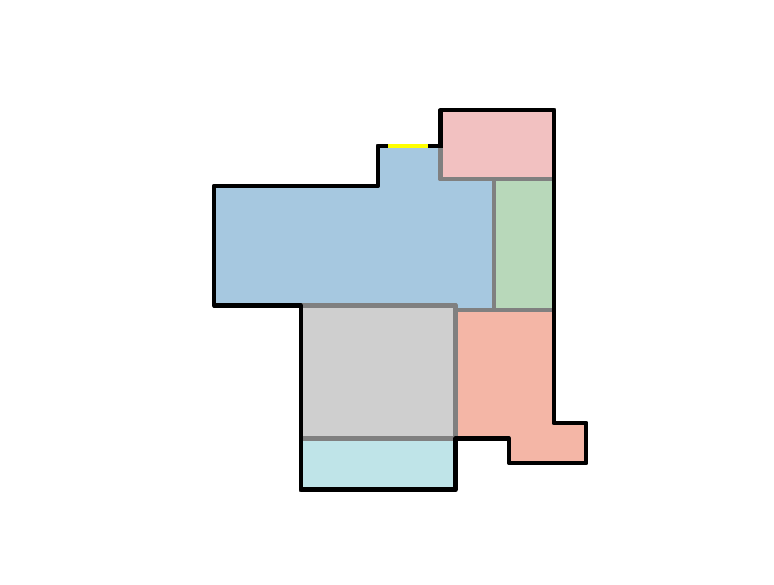} \\

    \end{tabular}
\end{minipage}

\begin{minipage}{1.0\textwidth}
\centering
    {\color{LivingRoomColor}\rule{2ex}{1.5ex}}~--~living room,
    {\color{MasterRoomColor}\rule{2ex}{1.5ex}}~--~master room,
    {\color{BathroomColor}\rule{2ex}{1.5ex}}~--~bathroom,
    {\color{KitchenColor}\rule{2ex}{1.5ex}}~--~kitchen,
    {\color{StudyRoomColor}\rule{2ex}{1.5ex}}~--~study room,
    {\color{SecondRoomColor}\rule{2ex}{1.5ex}}~--~second room,
    {\color{BalconyColor}\rule{2ex}{1.5ex}}~--~balcony
\end{minipage}

\caption{\label{fig:qualitative_comparison}Qualitative comparison of results between baseline, our method and samples from RPLAN dataset. Reduction of inaccessible bathrooms and blind corridors compared to the baseline; the final column shows improved room arrangements.}

\end{figure*}

\end{document}
}

\section{Results}
\vspace{-2mm}

We train on RPLAN~\cite{RPLAN} (quantized to 256). After validity filtering, the dataset contains 80,405 plans. We use a 90/5/5\% train/val/test split. The training set was expanded to $\sim$159,000 examples via augmentation (rotation, symmetry, room permutation).
The model (GPT-2, 25 layers, 16 heads, 256 embedding dim, 320 token context, 20M parameters) is trained for 150 epochs.
A baseline model with identical configuration but trained with standard cross-entropy loss \del{(no ergonomic term)} serves as a baseline. 
Inference uses greedy decoding. During testing the model starts from ground truth boundaries and doors (as constraints), though from-scratch generation is also possible.
\rev{Code and additional results are available on \href{https://comfortableworld.github.io}{https://comfortableworld.github.io}}.

\textbf{Qualitative results.} Figures~\ref{fig:teaser} and \ref{fig:qualitative_comparison} illustrate our results. Ergonomic loss minimization is evident in bathroom placement (closer to master/secondary rooms) and reduced blocking of functional rooms (e.g., kitchen blocking bathroom in baseline, Fig.~\ref{fig:teaser}).

\textbf{Quantitative results.} We define test metrics: \textit{Parsability} (successfully parsed sequences); \textit{Validity} (no self-intersections); \textit{Fully covered} (complete interior coverage); \textit{No room overlapping}; \textit{Ergonomic cost} (\rev{see Eq.~\ref{eq: ergonomic cost},} \del{Sec.~\ref{sec:ergo cost},} lower is better); and \textit{Perfect ergonomic cost} (\% with 0 cost).
\rev{All metrics, besides ergonomic cost are binary.}
Table~\ref{tab:results} compares these metrics. Our model outperforms the baseline on ergonomics while maintaining comparable generation quality. \del{The dynamic loss weighting (Sec.~\ref{sec:loss_usage}) enables learning ergonomic layouts despite the imperfect ergonomics\del{ adherence} of the training data.}

\documentclass[ShortPaper]{subfiles}

\begin{document}

\begin{table}[b]
\small
    \renewcommand{\arraystretch}{1.0}    \caption{\label{tab:results}Quality metric values for test results of both the baseline and our full method as well as ground truth RPLAN dataset metrics.}
    \centering
    \begin{tabular}{|c|c|c|c|}
        \hline
        \textbf{Metrics}             & \textbf{Baseline} & \textbf{Our}       & \textbf{RPLAN} \\ \hline
        
        Parsability (\%)             & \underline{99.93} & \underline{99.93}  & 100              \\ \hline      
        Validity (\%)                & 96.98             & \underline{97.71}  & 100              \\ \hline
        Fully covered (\%)           & \underline{79.83} & 74.19              & 100              \\ \hline
        No overlapping (\%)          & \underline{98.02}             & 97.76  & 100              \\ \hline
        Ergonomic cost (m)           & 0.502             & \underline{ 0.353} & 0.509          \\ \hline
        Perfect ergonomic cost (\%)  & 34.32             & \underline{43.75}  & 28.07          \\ \hline
    \end{tabular}
\end{table}

\end{document}

\section{Limitations and Conclusions}
\vspace{-2mm}

\textbf{Limitations.} While our method successfully optimizes ergonomics, we observe a trade-off in geometric packing efficiency, resulting in slightly lower area coverage compared to the baseline. Additionally, the current model operates unconditionally, lacking explicit constraints for room counts or types\del{, which limits direct user control}. Finally, rare logical inconsistencies persist, such as entrance doors opening into \del{unconventional} spaces like balconies.

\textbf{Conclusions.} We presented a transformer-based \del{data-driven}method for residential floor plan generation that integrates domain-specific ergonomic principles. By incorporating differentiable loss functions for room adjacency\del{ and proximity}, our model outperforms the baseline in ergonomic metrics despite the limitations of the training data. This approach demonstrates that incorporating domain-specific expert knowledge into training can effectively guide generative models toward more architecturally compliant and practical designs. 

\del{As future work, we plan to add conditional control over room types and counts and to improve geometric packing and logical consistency while preserving the benefits of the guided loss.}

\rev{As future work, we plan to add conditional control over room types and counts, improve geometric packing and logical consistency, and experiment with bigger, more recent model architectures.}

\begin{revision}
\section{Acknowledgments}
\vspace{-2mm}

This research was carried out with the support of the High Performance Computing Center at Faculty of Mathematics and Information Science Warsaw University of Technology.

Generative AI tools were used for brainstorming and polishing the manuscript text. All scientific content and analysis were produced entirely by the authors.
\end{revision}

\bibliographystyle{eg-alpha-doi}
\bibliography{ShortPaper}

\end{document}